# Strain-Induced Reversible Manipulation of Orbital Magnetic Moments in Ni/Cu Multilayers on Ferroelectric BaTiO$_3$

Jun Okabayashi[1*], Yoshio Miura[2], and Tomoyasu Taniyama[3]

[1]*Research Center for Spectrochemistry, The University of Tokyo, Bunkyo-ku, Tokyo 113-0033, Japan*
[2]*Research Center for Magnetic and Spintronic Materials, National Institute for Materials Science (NIMS), Tsukuba 305-0047, Japan*
[3]*Department of Physics, Nagoya University, Furo-cho, Chikusa-ku, Nagoya 464-8602, Japan*

(2019.4.12)

*jun@chem.s.u-tokyo.ac.jp

## ABSTRACT

Controlling magnetic anisotropy by orbital magnetic moments related to interfacial strains has considerable potential for the development of future devices using spins and orbitals. For the fundamental physics, the relationship between strain and orbital magnetic moment is still unknown, because there are few tools to probe changes of orbital magnetic moment. In this study, we developed an electric-field- ($E$)-induced X-ray magnetic circular dichroism (EXMCD) technique to apply $E$ to a ferroelectric $BaTiO_3$ substrate. We reversibly tuned the interfacial lattice constants of Ni/Cu multilayers on $BaTiO_3$ using this technique. As the domain structures in $BaTiO_3$ are modulated by $E$, EXMCD measurements reveal that the changes in the magnetic anisotropy of Ni/Cu films are induced through the modulation of orbital magnetic moments in Ni with magneto-elastic contributions. The strained Ni layer that induces the perpendicular magnetic anisotropy without $E$ is released at $E$ = 8 kV/cm, and in-plane magnetization also occurs. We observed that EXMCD measurements clarified the origin of the reversible changes in perpendicular magnetic anisotropy and established the relationship between macroscopic inverse magnetostriction effects and microscopic orbital moment anisotropy.

## Introduction

The coupling between ferromagnetic and ferroelectric properties has recently attracted considerable attention toward the creation of novel devices using multiferroic controlling of their properties[1–7]. In particular, the hetero-interfaces in thin films comprising both ferromagnets and electrically polarized materials produce a rich variety of possibilities for creating multifunctional properties[8–18]. Modulation of interfacial lattice constants by an electric field ($E$) induces interfacial changes in magnetism. The interfacial lattice distortion produces variations in magnetic properties, which are recognized as inverse magnetostriction effects[19–23]. Moreover, magnetic anisotropy is tuned by lattice distortions. Recently, the magnetic anisotropy controlled by $E$ has become an important subject in spintronics, which is the study aiming at the realization of devices operating with low energy consumption[24–28]. Recent attempts have been focused on the modulation of the number of charge carriers at the interface between an ultrathin ferromagnetic layer and an oxide-barrier insulator in magnetic tunnel junctions. Other approaches, which are our focus in this study, are based on the interfacial mechanical-strain coupling between ferromagnetic and ferroelectric layers using multiferroic hybrid structures. As one of the candidate approaches, applying $E$ to $BaTiO_3$ provides the possibility to tune the lattice constants by modulating the domain structures along the $a$- and $c$-axes directions by 3.992 Å and 4.036 Å, respectively, at room temperature[29,30].

There exist some reports related to depositing magnetic thin-film layers onto $BaTiO_3$ for $E$-induced magnetism with abrupt interfaces between $BaTiO_3$ and Fe or Co[12,14–16,21,22,24]. In other cases, thin Ni layers sandwiched by Cu layers exhibit perpendicular magnetic anisotropy (PMA) because of the interfacial tensile strain in the Ni layers[31]. Recently, $E$-control of the magnetic properties of Ni/Cu multilayers on $BaTiO_3$ was achieved; the magnetization was switched from the perpendicular axis to the in-plane easy axis by tuning the lattice distortion through the application of $E$[32]. These effects may be explained phenomenologically by inverse magnetostriction effects. Although anisotropic energies depend on orbital magnetic moments, the microscopic origin of the control of the anisotropic energy dependent on strain is still not known explicitly. Theoretical approaches that consider spin–orbit interactions, as well as crystalline potentials, as perturbative treatments have been developed[33]. Moreover, strain-induced orbital magnetic moments have been discussed alongside calculations on the strained Ni layers[34], but the relationship between strain and orbital moments is still unknown.

Element-specific magnetic properties and their origins should be investigated by applying $E$ explicitly to clarify the relationship between lattice distortion and magnetic properties. In particular, magnetic anisotropy is related to the anisotropy of orbital magnetic moments[35]. A unique tool to deduce spin moments as well as orbital moments is X-ray magnetic circular dichroism (XMCD) with magneto-optical sum rules[36,37]. Recent developments of XMCD by applying $E$ have focused on charge accumulation at the interface of FePt/MgO[38], interfacial oxidation reaction of Co/Gd$_2$O$_3$[39], and other cases[40–42]. In the case of FePt/MgO, the difference in the Fe XMCD by applying $E$ is negligibly small because of the quite small amounts of charge accumulation at the interfaces. In the case of Co/Gd$_2$O$_3$, the chemical reaction at the oxide interfaces becomes dominant. Investigating the orbital moment anisotropy (OMA) by applying $E$ is a challenging approach to initiate novel research into the physics of the relationship between lattice distortion and orbital magnetic moments, which is a fundamental and unsolved problem in the scientific research field. Considering the relationships between spin magnetic moments ($m_s$), orbital magnetic moments ($m_{orb}$), and strain ($\varepsilon$), the spin–orbit interaction links $m_s$ and $m_{orb}$, and the magnetostriction links $m_s$ and $\varepsilon$. However, the relationship between $m_{orb}$ and distortion is still unexplored. To understand this relationship and the elastic phenomena from the view point of $m_{orb}$, we developed a technique by applying an electric field in XMCD measurements (EXMCD) to clarify the mechanism of the electric-field-induced changes in the magnetic anisotropy of Ni/Cu multilayers on BaTiO$_3$ hetero-structures through lattice distortions. In this study, we aim to clarify the relationship between strain and orbital magnetic moments using the EXMCD method; we also present first-principles calculation of the changes in magnetic anisotropy.

## Results

**Strain introduced into the samples**

First, we mention the values of the strain introduced into the samples, as illustrated in Fig. 1a. Without applying $E$, the Ni layer possesses a tensile strain of 2% through the sandwiched-Cu layers, and the Ni layer exhibits PMA as shown in Fig. 1b. When $E$ is zero, the $a$- and $c$-domain structures are mixed in BaTiO$_3$. By applying $E$, the $c$-domain structures become dominant, from which it may be inferred that the application of the electric field, $E$, compresses the lattice constant of BaTiO$_3$ and releases the strain in the Ni layer, thereby resulting in the magnetization in the in-plane easy axis in the Ni layers. Therefore, in the $a$- and $c$-domain structures of BaTiO$_3$, the Ni layer exhibit PMA and in-plane anisotropy, respectively[32]. Figure 1c displays the differential interference microscopy images of the $a$- and $c$-domains before and during the application of $E$. The bright area indicates the $a$-domain structure. Both $a$- and $c$-domain structures without $E$ are clearly observed, and they align to the $c$-domain structure by applying ±5 kV/cm, which is consistent with the magnetization. The area of $a$-domain structure is estimated from the images in Fig. 1c as approximately 50 % without $E$. We emphasize that the strain propagation proceeds into all multilayers by the modulation of lattice constants of BaTiO$_3$ substrates due to the high strain-transfer parameter value.[32,43] In the case of 20-nm-thick Fe film on BaTiO$_3$, strain propagation into the surfaces is also detected.[44] Further, we confirmed that this process is reversible by removing $E$.

**XAS and XMCD**

Results from X-ray absorption spectroscopy (XAS) and XMCD in the total-electron-yield (TEY) mode, which probe the depth beneath 3 nm from the surface by collecting drain currents of secondary photoelectrons, without applying $E$, are shown in Fig. 2. The intensity ratio between the $L$-edges of Ni and Cu guarantees the amounts of Ni and Cu within the detection limits in TEY. We observed clear XMCD at the $L$-edge of Ni but not at the $L$-edge of Cu. This phenomenon suggests that the magnetic moments at the interface are not induced into the Cu layers. Considering the magneto-optical sum rules[36,37], the spin and orbital magnetic moments in the Ni sites are estimated to be 0.50 and 0.04 $\mu_B$, respectively, by assuming that the hole number is 1.75. These values are comparable to previous reports of the Ni/Cu interface[44]. Angular-dependent XMCD measurements deduce the spin dipole term $m_T$ of smaller than 0.005 $\mu_B$.

Figure 3 shows the Ni $L$-edge XAS and XMCD spectra obtained in the partial fluorescence yield (PFY) mode by applying an electric field, $E$, of positive bias of 8 kV/cm. The PFY mode probes a depth beneath approximately 100 nm from the surface because of the photon-in and photon-out processes. Sample surfaces are connected to ground and $E$ is applied to the back side of the BaTiO$_3$ substrates. The spectral line shapes of the XAS and XMCD are modulated by $E$ in normal incident (NI) case in spite of fixed sample measurement position. By comparing the results with and without $E$, we observed a slight variation in the peak asymmetries between the $L_3$ and $L_2$ edges. Moreover, the integrals of the $L_{2,3}$-edge XMCD peaks are plotted on the same panel to emphasize the difference in the spectral line shapes with the application of $E$. As the convergent values of the integrals of the XMCD are proportional to the orbital magnetic moments within the framework of the XMCD sum rule[36,37], these results reveal that the orbital moments are modulated by applying $E$, thereby resulting in changes of magnetic anisotropy. As the vertical beam size is approximately 1 mm, the contribution from only the *a*-domain cannot be detected in the 0 kV/cm condition, which underestimates the perpendicular component of the orbital moments. The values of spin and orbital moments are estimated to be 0.56 and 0.055 $\mu_B$, respectively, for an electric field of 0 kV/cm, and 0.56 and 0.045 $\mu_B$, respectively, for an electric field of 8 kV/cm with error bars of ±20% for each value considering estimated ambiguities applying sum rules as listed in Table I. The modulation of the orbital magnetic moments by 0.01 $\mu_B$ upon applying $E$ is related to the induced lattice distortion of 2% from the BaTiO$_3$ substrates. Moreover, after releasing $E$ to zero, spectral line shapes also revert to the pristine state. For grazing incident (GI) case, XAS and XMCD spectra with and without $E$ are displayed in Figs. 3c and 3d. The values of $m_s$ and $m_{orb}$ are also listed in Table I. Effects of electric field in GI are smaller than those in NI because oblique configuration of 60° from sample surface normal detects the half of in-plane components (cos60°=1/2). Angular dependence depicts the changes of $m_s^{eff}$ which includes $m_s+7m_T$ and the magnetic dipole term of $m_T$ cancels in the magic angle 53.7° of near GI set up. Details are explained in supplemental material. Thus, the value of $m_T$ is estimated less than 0.001 $\mu_B$. Therefore, these results originate from the modulation of not spin moments but orbital moments, suggesting that the inverse magneto-striction effects are derived from the changes of orbital moments.

The element-specific magnetization curves (*M*–*H* curves) at the $L_3$-edge of Ni during the application of $E$ in the normal incidence setup are shown in Fig. 4. As the normal of the sample surface is parallel to both incident beam and magnetic field, the contribution from the easy axis in PMA is observed. By applying an electric field, $E$, of ±8 kV/cm, the *M–H* curves change to those of the in-plane easy-axis behavior, which is related to the changes in the orbital magnetic moments in Ni. After switching off $E$, the *M–H* curves exhibit the PMA characteristics again, as shown in Fig. 4b. Moreover, the reversible changes observed by applying $E$ in XMCD are confirmed by the changes in the XMCD line shapes. The amounts of the changes in the *M–H* curves at the Ni $L_3$-edge XMCD are a little similar to those measured by the magneto-optical Kerr effect[32]. This phenomenon suggests that the domain structures in the observed area in the EXMCD measurements can be changed from the *a*-domain to the *c*-domain by applying $E$.

**First-principles density-functional-theory calculation**

We performed first-principles calculations of magneto-crystalline anisotropy (MCA) energies for fcc Ni as a function of the in-plane lattice constant ($a_\parallel$). Assuming the motion of free electrons as a ground state, spin-orbit interaction is adopted as a perturbation term for the estimation of MCA energy. The MCA energy is defined as the difference between the sums of the energy eigenvalues for magnetizations oriented along the in-plane [100] and out-of-plane [001] directions ($\Delta E_{MCA}=E[100]-E[001]$). We employed the spin–orbit coupling constant ξ of the Ni atom, 87.2 meV, in the second-order perturbation calculation of the spin–orbit interaction to obtain MCA energies for each atomic site.[46]

Figure 5a shows the $\Delta E_{MCA}$ and the anisotropy of orbital moments ($\Delta m_{orb}=m_{orb}[001]-m_{orb}[100]$) as a function of $a_\parallel$, where a perpendicular lattice parameter for each $a_\parallel$ is optimized from the equilibrium value of $a_0$ = 3.524 Å. As shown in Fig. 5a, both $\Delta E_{MCA}$ and $\Delta m_{orb}$ increase with the tensile in-plane distortions, which are consistent with the XMCD results. In the equilibrium condition in the fcc structure, orbital and spin moments of 0.0483 μ$_B$ and 0.625 μ$_B$, respectively, are estimated. This estimation is in good agreement with previous band-structure calculations for Ni[34,47]. The slope in Fig. 5a results in a modulation of the orbital moment of 0.002 μ$_B$ per 1% strain. We display in Fig. 5b the spin-resolved MCA energies for the four cases $\Delta E_{\uparrow\uparrow}$, $\Delta E_{\downarrow\downarrow}$, $\Delta E_{\uparrow\downarrow}$, and $\Delta E_{\downarrow\uparrow}$ as a function of $a_\parallel$, where ↑ and ↓ indicate majority(up)- and minority(down)-spin states, respectively; the left arrow indicates an initially occupied spin-state and the right arrow indicates an intermediate spin-state in the second order perturbation of the

spin–orbit interaction[46]. First, the spin-conservation term $\Delta E_{\downarrow\downarrow}$ increases with increasing $a_\parallel$ of fcc Ni and qualitatively reproduces the $a_\parallel$ dependence of the MCA energies and $\Delta m_{\text{orb}}$. Second, the spin-flip term $\Delta E_{\uparrow\downarrow}$ decreases with increasing $a_\parallel$, indicating that the origin of the change of the MCA energies of fcc Ni by the tetragonal distortion can be attributed to the $\Delta E_{\downarrow\downarrow}$. We confirmed the strain dependence of spin magnetic moments is ten times smaller than $\Delta m_{\text{orb}}$. Therefore, this means that the MCA of fcc Ni can be described mainly by Bruno's relation through the orbital moment anisotropy[35]. Since $\Delta E_{\downarrow\downarrow}$ and $\Delta E_{\uparrow\downarrow}$ can be described as the energy differences between the $z$ and $x$ directions,

$$\Delta E_{\downarrow\downarrow} = E_{\downarrow\downarrow}(x) - E_{\downarrow\downarrow}(z) = -\xi^2 \sum_{o\downarrow,u\downarrow} \frac{|\langle o\downarrow|L_x|u\downarrow\rangle|^2 - |\langle o\downarrow|L_z|u\downarrow\rangle|^2}{E_{u\downarrow} - E_{o\downarrow}}$$

$$\Delta E_{\uparrow\downarrow} = E_{\uparrow\downarrow}(x) - E_{\uparrow\downarrow}(z) = \xi^2 \sum_{o\uparrow,u\downarrow} \frac{|\langle o\uparrow|L_x|u\downarrow\rangle|^2 - |\langle o\uparrow|L_z|u\downarrow\rangle|^2}{E_{u\downarrow} - E_{o\uparrow}} \quad . \tag{1}$$

The matrix elements of $L_x$ and $L_z$, depending on in-plane strain, provide the orbital-resolved contributions; $o(u)$ represents occupied (unoccupied) states.[48] We adopted spin-orbit coupling constant $\xi$ of 78 meV for Ni. The matrix elements of $L_z$ and $L_x$ favor the out-of-plane and in-plane contributions, respectively. The matrix elements of $L_z$ between $d(xy)$ and $d(x^2\text{-}y^2)$ orbitals are large positive contributions to the MCA energies and increase with $a_\parallel$. Each element is estimated as a function of the strain, as shown in Fig. S2 (See supplemental material). Thus, the changes of orbital hybridization in $d(xy)$ and $d(x^2\text{-}y^2)$ orbitals directly contribute to the change of the MCA energy in $\Delta E_{\downarrow\downarrow}$ through the tetragonal distortions. These pictures appear in the band dispersions of strained fcc Ni. Figures 5c and 5d show the minority band dispersion of fcc Ni without distortion and with 2% extensive distortion. The color map of the band dispersion indicates the magnitude of the projection of the $d(x^2\text{-}y^2)$ and $d(xy)$ orbitals. As shown in Figs. 5c and 5d, the $d(x^2\text{-}y^2)$ bands around the M point approach the Fermi energy, while the $d(xy)$ bands move very little in response to the tensile distortion. Since the $d(x^2\text{-}y^2)$ orbital spreads in the direction of the nearest neighbor atoms, it is strongly affected by in-plane distortion; whereas, the influence of the distortion is relatively small for the $d(xy)$ orbital because of the distribution between the nearest neighbor atoms. Therefore, first-principles calculations also capture the trends of the modulation in the $d(xy)$- and $d(x^2\text{-}y^2)$-orbital states. Furthermore, the electron occupation number in Ni is estimated to be 8.25 which remains unchanged by the introduction of strain because of the compensation of the occupancy dependence of each 3*d* orbital.

## Discussion

Considering the above results, we discuss the relationship between the OMA and the magneto-elastic energy. In particular, we analyze the magnetic anisotropy energies in the strained Ni layers depending on the magnetization and the lattice distortion. Microscopically, the OMA can be described by Bruno's relation[35] through the second-order perturbation of the spin–orbit interaction. Moreover, the OMA produces the crystalline anisotropy $\Delta K = \alpha \xi \Delta m_{orb}$, where $\Delta m_{orb}$ is the difference between the orbital moment of the component perpendicular to the film and that of in-plane component, with the coefficient of the spin–orbit coupling constant $\xi$ and the band-structure parameter $\alpha=1/4$ for a more-than-half-filled 3*d* transition metal Ni. The XMCD shown in Fig. 3 clearly exhibits OMA that depends on the applied electric fields. The value of $\Delta m_{orb}$ is estimated to be 0.01 $\mu_B$ for a strain modulation of 2%, which results in the anisotropy energy of $6.8 \times 10^5$ J/m$^3$ by assuming fcc-Ni lattice constants of 3.524 Å. As the hysteresis curves in Fig. 1 and the EXMCD results in Fig. 4 are almost identical, similar anisotropy energies can be obtained. The first-principles calculation also reproduces the $\Delta K$ of the order of $10^5$ J/m$^3$. The anisotropy energy $\Delta K$ is formulated in the scheme of OMA by including the magneto-elastic energy as a function of strain ($\varepsilon$);

$$\Delta K = \alpha \xi \Delta m_{orb}(\varepsilon) = \alpha \xi 0.005 \varepsilon . \qquad (2)$$

As the interfacial strain modulates the orbital magnetic moment, the strength of the distortion $\varepsilon = \Delta l / l$, quantified as a ratio of the length difference, is scaled to $\Delta m_{orb}$, which is deduced from the band-structure calculation. In general, eq. (2) consists of two terms of $\Delta m_{orb}$ and $m_T$. However, the contribution of $m_T$ is much smaller than $\Delta m_{orb}$ in 3*d* TMs. Then, we focus on only the first term. Figure 5a confirms the linear relationship between $\varepsilon$ and $\Delta m_{orb}$. The slope in Fig. 5a quantitatively describes the dependence of $\Delta m_{orb}$ on $\varepsilon$ in Eq. (2). By applying $E$ to BaTiO$_3$, the released lattices shorten the in-plane lattice distances, thereby resulting in the decrease of perpendicular orbital moments. Quantitatively, the 2% modulation of the lattice generates the OMA of 0.01 $\mu_B$, which is of the same order as that deduced from the first-principles calculations.

Next, we discuss the discrepancy between XMCD and first-principles calculation. The discrepancy might be understood as the underestimating of $m_{orb}$ in the calculation because of the lack of considering Hund's second rule in the electron correlation[49]. By considering orbital polarization, $\Delta m_{orb}$ estimated from the first-principles calculation becomes similar to that deduced from XMCD. Other reasons for the discrepancy might originate in the spin-flipped contribution through $\Delta E_{\downarrow\uparrow}$. On the other hand, EXMCD measurements also deduced the magnetic dipole term in Ni, the order of which is smaller than $\Delta m_{orb}$. However, this term proposed by van der Laan[50] which is deduced from $\Delta E_{\uparrow\downarrow}$ in eq. (1) is smaller than OMA. Therefore, the modulation of the magnetic anisotropy introduced by the macroscopic strain in the Ni layers is connected mainly with the OMA as a microscopic origin. Further, EXMCD and first-principles calculation explain qualitatively that the $m_s$ values are less sensitive to the strain and orbital hybridization, resulting in the changes of $m_{orb}$.

In conclusion, by using the novel EXMCD technique, we clarified that the reversible PMA changes in the Ni/Cu film on $BaTiO_3$ are induced by the modulation of orbital magnetic moments in Ni. The strained Ni layer that induces the PMA without $E$ is released upon the application of an $E$-field and is modulated to produce in-plane magnetization. Moreover, the magnetization curves in the Ni $L_3$-edge EXMCD measurements are modulated between out-of-plane and in-plane magnetization. We revealed that the changes of magnetic anisotropy by $E$, which were explained by the phenomenological magneto-elastic description, may be understood microscopically by OMA. These results introduce the concept of orbital-striction or orbital-elastic effects at the hetero-interfaces beyond established magnetostriction effects.

## Methods

**Sample preparation.** The samples were grown by using ultra-high vacuum molecular beam epitaxy on [100]-oriented 0.5-mm-thick $BaTiO_3$ single crystal substrates. Therefore, the bias voltage of 400 V applied between top and bottom electrodes means the electric field of 8 kV/cm. The stacked structures are shown in Fig. 1a. Before the deposition of face-centered-cubic (fcc) Ni/Cu stacked multilayers, a 1-nm-thick Fe buffer layer was deposited onto the substrate at 300 °C. The multilayers of [Cu (9 nm)/Ni (2 nm)]$_5$ were grown at room temperature and covered by 1-nm-thick Au to prevent oxidization. The details of the surface and interface conditions and the fabrication procedures are reported in Ref. 32. The magnetic properties were characterized by magneto-optical Kerr effect (MOKE) measurements and magnetometry. The ferroelectric domain structures were observed by differential interference microscopy before the EXMCD measurements.

**XAS and XMCD measurements.** The XAS and XMCD measurements for the Ni and Cu *L*-edges were performed at the KEK-PF BL-7A beamline, Japan, at room temperature. Magnetic fields of ±1.2 T were applied along the incident polarized soft X-rays to saturate the magnetization sufficiently along the normal direction of the surface of the sample. The TEY mode was adopted by detecting the drain currents from the samples for the case of the measurements recorded without applying a magnetic field. The electrodes were mounted at the surface of the sample and at the rear of the substrate to perform the EXMCD measurements. The EXMCD measurements were performed using the PFY mode to probe the signals beneath more than 10 nm below the surfaces of the samples using a bipolar electric power source (Keithley 2410) to apply *E* to BaTiO$_3$. The fluorescence signals were detected by a silicon drift detector (Princeton Gamma-Tech. Instrument Inc. SD10129), mounted at 90° to the incident beam. The XAS and XMCD measurements were performed in the normal-incidence setup, in which the normal of the sample's surface is parallel to the incident beam and the magnetic field, detecting the signals of perpendicular components to the films. We changed the magnetic field directions to obtain the right- and left-hand-side polarized X-rays while fixing the polarization direction of the incident X-ray. To avoid saturation effects in the PFY mode, the intensities in the XAS measurements were carefully examined by comparing them with those obtained in the TEY mode. Analysis method of $m_{orb}$ and $m_s$ using sum rules is described in Supplemental material and ref. 51.

**First-principles study.** The DFT calculation code of the Vienna ab initio simulation package (VASP), including the spin–orbit interaction with the spin-polarized generalized gradient approximation, was employed[52-54]. The plane-wave cutoff energy was set to 500 eV and a 25 × 25 × 17 *k*-point mesh was used for sampling the Brillouin zone. The coordinate system of the Ni crystal lattice used in the calculation is as follows. The *z*-axis is the *c*-axis direction (perpendicular magnetization direction), and the *x*-axis is the direction of the nearest neighbor atom in the in-plane direction. With the changing of the in-plane lattice constant, the *c*-axis length was optimized in the first-principles calculations.

## Acknowledgments

This work was partly supported by JSPS KAKENHI (Grant Nos. 15H03562, 16H06332, 15H01998, 16K14381, and 17H03377), Spintronics Research Network of Japan, JST CREST Grant Number JPMJCR18J1, Japan, and the Asahi Glass Foundation. The synchrotron radiation experiments were performed under the approval of the Photon Factory Program Advisory Committee, KEK (No. 2017G060). The authors acknowledge Dr. Seiji Mitani for fruitful discussion.

## Author Contributions

J.O. and T.T. planned the study. T.T. prepared the samples and performed the magnetization measurements. J.O. set up the EXMCD measurement apparatus at Photon Factory and collected and analyzed the data. Y.M. performed the first-principles calculation. All authors discussed the results and wrote the manuscript.

## Additional Information

**Supplemental information** accompanies the paper on the npj Quantum Materials website (http://***).

## Competing interests

The authors declare no Competing Financial or Non-Financial Interests.

## Data availability

The data that support the findings of this study are available from the corresponding author upon reasonable request.

# Figure Legends

**Figure 1 | Sample growth and characterization by applying electric fields.** (**a**) Schematic diagram of the structure of the sample indicating with film thicknesses indicated. The electrodes for applying the electric field are mounted at the top (Ni/Cu layer) and the bottom (BaTiO$_3$ substrate) of the films. (**b**) *M–H* loops measured by vibrational sample magnetometer (VSM) in the easy- and hard-axes directions, without an applied electric field. (**c**) Top-view differential interference microscope images recorded before and during the application of the electric field. The white area indicates the *a*-domain that exhibits PMA in the Ni layers.

**Figure 2 | XAS and XMCD spectra obtained at Ni and Cu *L*-edges in total-electron-yield mode without an applied electric field.** The spectra for NiCu/BaTiO$_3$ were measured in the normal-incident geometry. (**a**) XAS results with differently polarized X-rays σ$^+$ (red) and σ$^-$ (blue). (**b**) XMCD result obtained from the difference of the σ$^+$ and σ$^-$ XAS spectra.

**Figure 3 | XAS and XMCD spectra obtained at Ni *L*-edges in partial-fluorescence-yield mode with and without an applied electric field.** The spectra for NiCu/BaTiO$_3$ were measured in the normal-incident geometry (**a**) and (**b**), in the grazing-incident geometry (**c**) and (**d**). Spectra recorded without an applied field *E* ((**a**), (**c**)) and with *E* of 8 kV/cm ((**b**), (**d**)); XAS with differently polarized X-rays (σ$^+$ and σ$^-$ are indicated by the red and blue lines respectively); XMCD obtained as the difference between the σ$^+$ and σ$^-$ XAS spectra. The integrals of the XAS and XMCD spectra are also shown in the same panel with their axis indicated on the right. Note that the scales of vertical axis are fixed in all panels. Illustrations in the XMCD panel show the schematic view of applying electric field and the angle between surface normal and incident beam direction.

**Figure 4 | Element-specific magnetization curves for the Ni *L*$_3$-edge of XMCD spectra with applied electric field.** The photon energy was fixed at the Ni *L*$_3$-edge in the PFY mode. Normal incidence set up was adopted. (**a**) Increasing the electric field from 0 kV/cm up to 8 kV/cm, and (**b**) decreasing the electric field from 8 kV/cm to 0 kV/cm.

**Figure 5 | First-principles calculations.** (**a**) MCA energies and anisotropy of orbital moments of fcc Ni as a function of the in-plane lattice constant $a_\parallel$ (tetragonal distortion is defined as $(a_\parallel - a_0)/(a_0 \times 100)$, where $a_0 = 3.524$ Å. (**b**) The second-order perturbative contribution of the spin–orbit interaction to the MCA energy of fcc Ni as a function of $a_\parallel$. The minority band dispersions of fcc Ni along the high-symmetry line without (left) and with (right) 2% tetragonal distortion for the (**c**) $d(x^2-y^2)$ and (**d**) $d(xy)$ orbitals. The color maps of the band dispersion indicate the magnitudes of projection of the $d(x^2-y^2)$ and $d(xy)$ orbitals.

Table I, The spin and orbital magnetic moments with and without $E$. The values are in the units of $\mu_B$. Experimental error bars are estimated about 20 % for the applications of XMCD sum rules.

|  | 0 kV/cm | 8 kV/cm |
| --- | --- | --- |
| $m_s$ (NI) | 0.56 | 0.56 |
| $m_{orb}$ (NI) | 0.06 | 0.04 |
| $m_s$ (GI) | 0.54 | 0.55 |
| $m_{orb}$ (GI) | 0.04 | 0.05 |

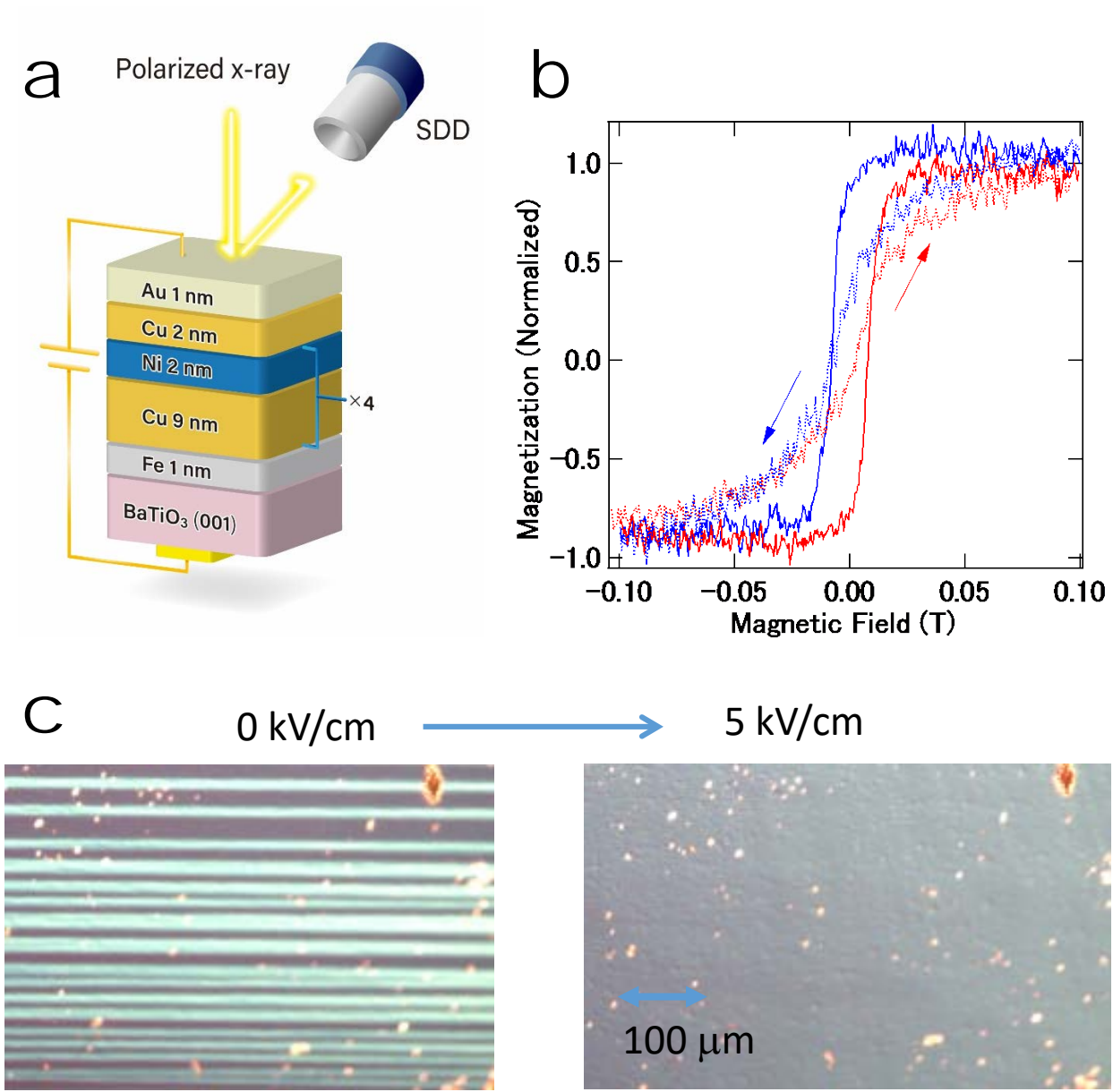

Fig. 1

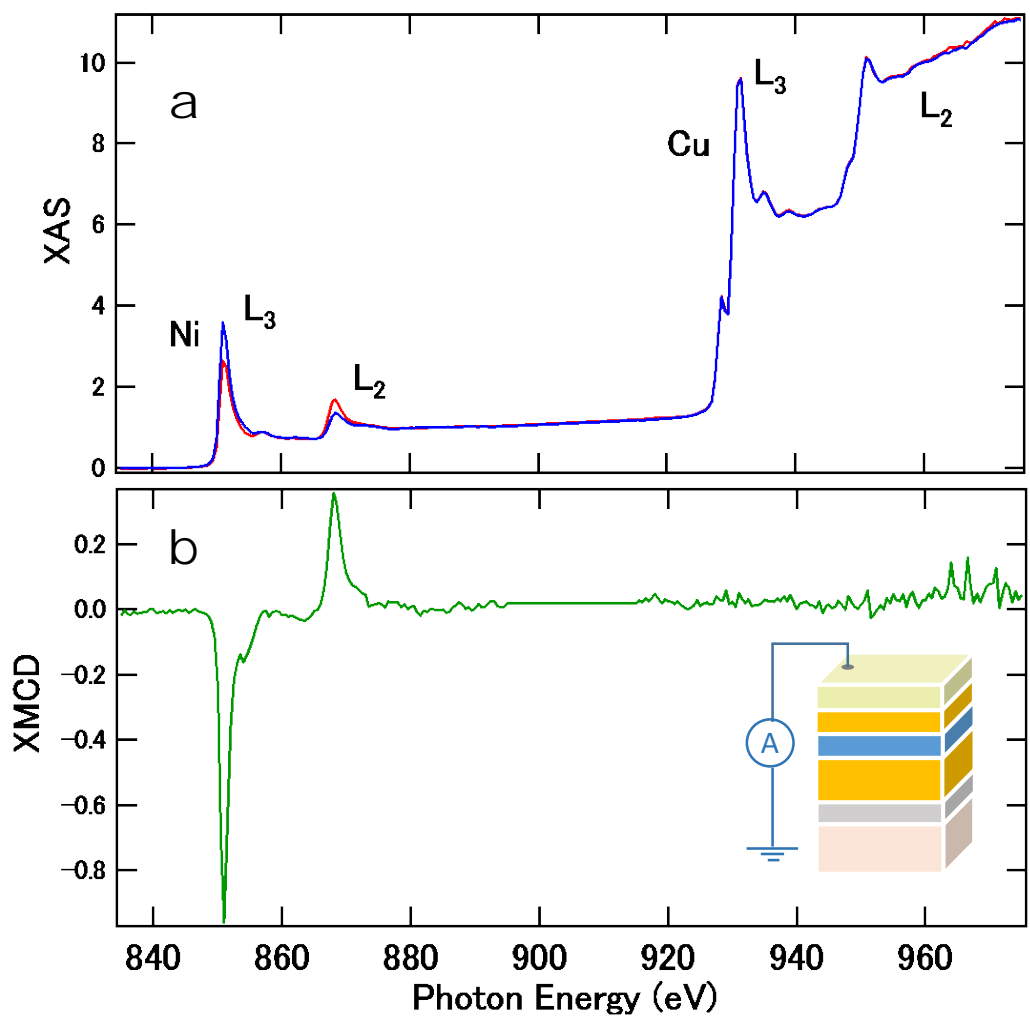

Fig. 2

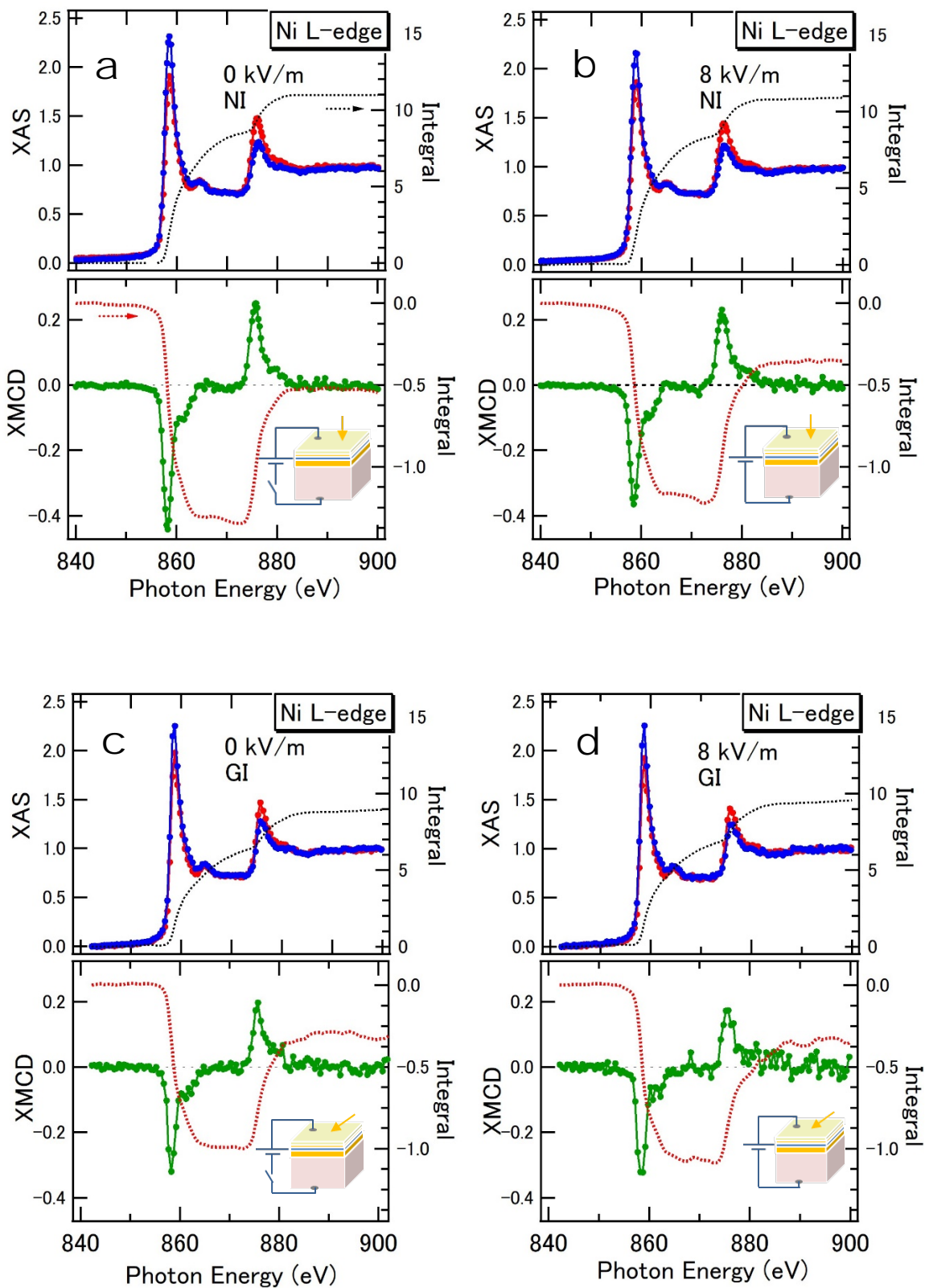

Fig. 3

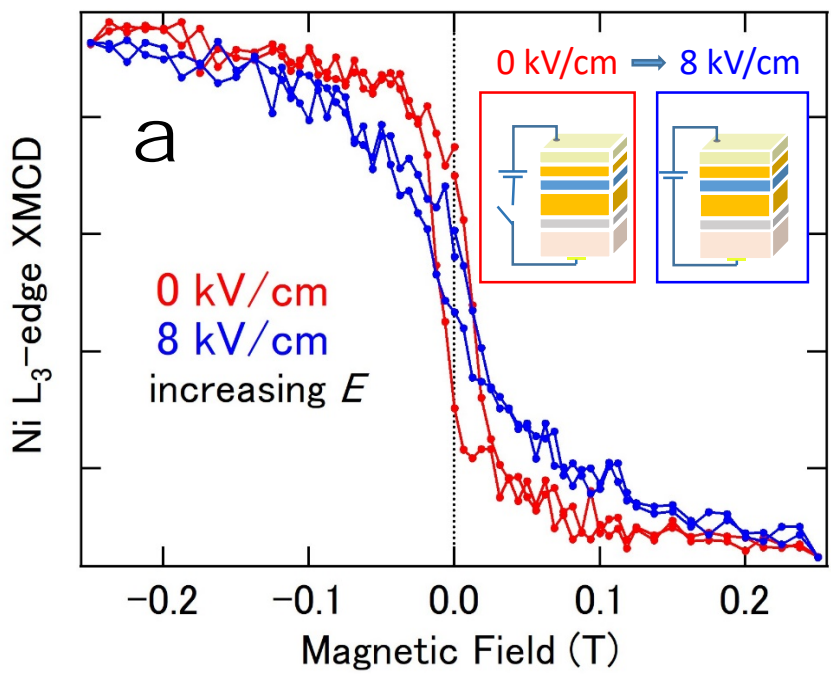

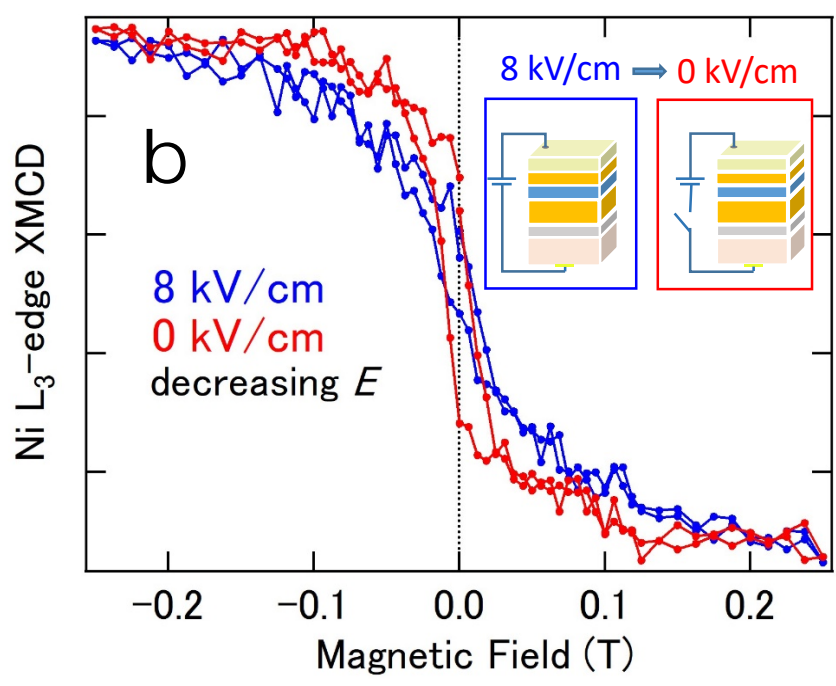

Fig. 4

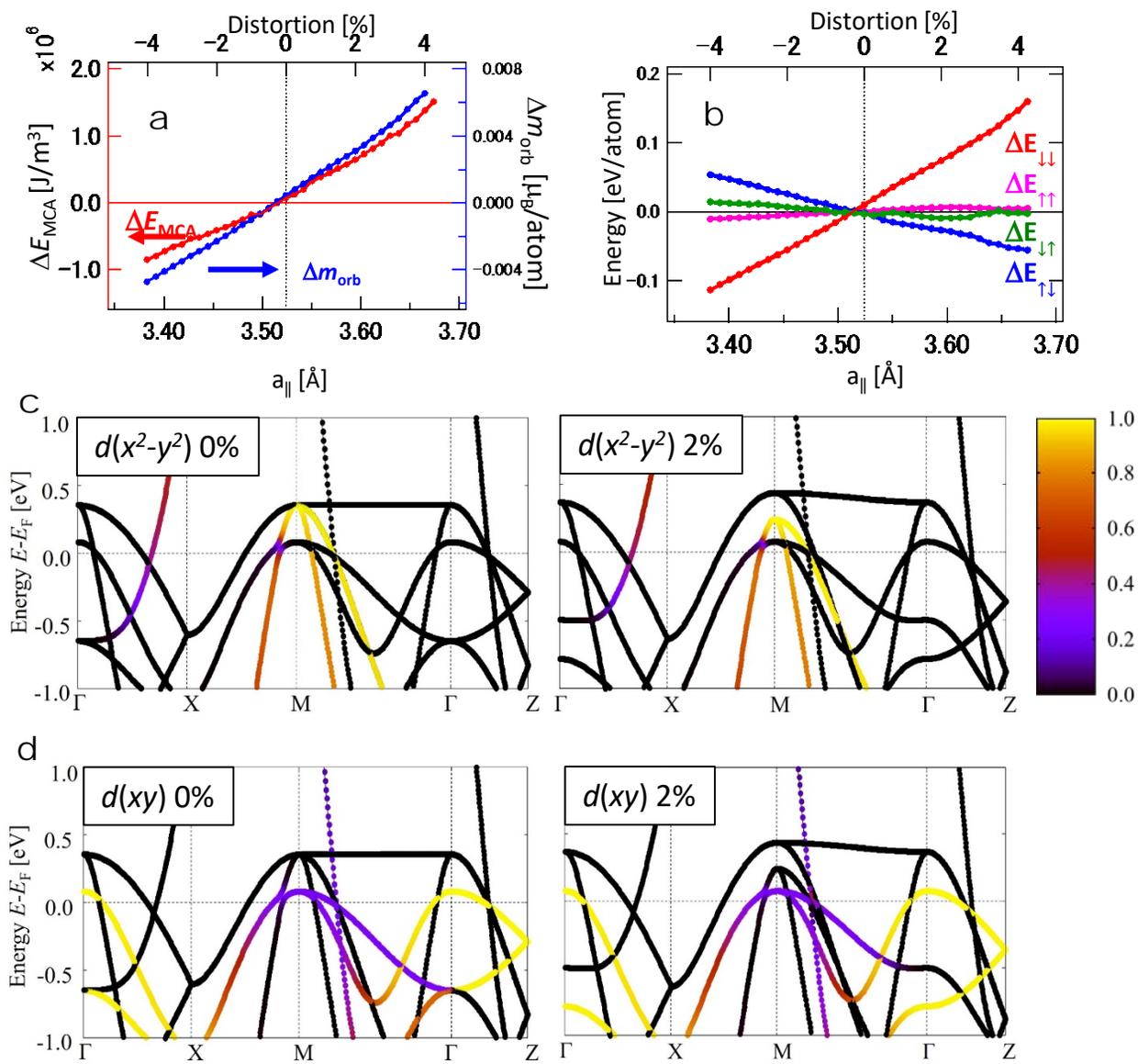

Fig. 5